\documentstyle[preprint,aps]{revtex}
\begin{document}
\draft
\preprint{SHS-96-7}

\def \beq{\begin{equation}}
\def \eeq{\end{equation}}
\def \beqarr{\begin{eqnarray}}
\def \eeqarr{\end{eqnarray}}


\newcommand{\half}{\frac 1 2 }
\newcommand{\eg}{{\em e.g.} }
\newcommand{\ie}{{\em i.e.} }
\newcommand{\be}{\begin{eqnarray}}
\newcommand{\ee}{\end{eqnarray}}
\newcommand{\noi}{\noindent}
\newcommand{\etal}{{\em et.al.\ }}

\newcommand{\dd}[2]{{\rmd{#1}\over\rmd{#2}}}
\newcommand{\pdd}[2]{{\partial{#1}\over\partial{#2}}}
\newcommand{\pa}[1]{\partial_{#1}}

\def\phidag{\phi^{\dagger}}
\def\fii{\varphi}
\def\fiis{varphi^\star}
\def\chidag{\chi^{\dagger}}
\def\chistar{\overline\chi}
\def\inta{\int_0^{2\pi}}
\def\aeo{a_1^0}
\def\ato{a_2^0}
\def\aev{\vec a_1}
\def\atv{\vec a_2}
\def\sz{\sigma_z}
\def\nh{\hat n}
\newcommand{\nhi}[1]{\hat n_{#1} }
\def\at{\tilde a}
\def\atvidv{\vec{\tilde a}}
\newcommand{\ki}[2]{l_{#1 #2}}
\newcommand{\km}[2]{K_{#1 #2}}
\def\rhobar{\bar\rho}
\def\ar{\alpha(r)}

\title{A Non-linear $\sigma$-model for Partially Polarized QH-states}
\author{ T.H. Hansson$^{1,3}$, A. Karlhede$^{1,3}$ and
 J.M. Leinaas$^{2,3}$}

\address{
$^1$Department of Physics,
Stockholm University,
Box 6730, S-11385 Stockholm,
Sweden
}

\address{
$^2$Institute of Physics,
University of Oslo,
P.O. Box 1048 Blindern, N-0316 Oslo,
Norway
}

\address{
$^3$Centre for Advanced Study,
P.O. Box 7606 Skillebekk, 0205 Oslo,
Norway
}

\date{June 17, 1996}
\maketitle
\begin{abstract}

We consider a two-component quantum Hall system within a
Landau-Ginzburg  theory with two Chern-Simons gauge fields. From this
theory we derive a  sigma model covariantly coupled to one Chern-Simons
field and find mean field solutions  that could describe partially
polarized quantum Hall states.   The quasiparticles in the original
model, which have quantized charge and spin, are described in the
covariant sigma model by topological  excitations, with the correct
quantum numbers. They  have finite energy due to the presence of the
Chern-Simons field, and closely resemble the skyrmions in the usual
non-linear sigma model. For the fully polarized states the spin is no
longer quantized, but determined by Coulomb and Zeeman interactions.

\end{abstract}
\pacs{}

There is a continuing strong interest in the Quantum Hall (QH) effect in
multi component systems \cite{gmreview}. The components can be electrons
with
 different spin, or electrons in different valleys (in Si systems), or
in different layers in multilayer systems. We refer to all such
components as spin.  A new feature of these
 systems is that there are quasiparticle excitations with  a
topologically nontrivial texturing of the components. The first  example
of this was the ``skyrmions'' involving the spin \cite{sondhi1,lee};
there is now strong experimental evidence that they are the lowest
energy quasiparticles in the fully polarized $\nu=1$ QH effect
\cite{barrett1,schmeller,goldberg}. Subsequently, a similar topological
excitation, the meron,  was used to  explain a novel phase transition
observed in double layer systems
\cite{indiana}. The textures have been studied  within a two dimensional
non-linear sigma model, describing the long distance physics, where the
quasi particles are the (baby) skyrmions,  as well as  within a
Hartree-Fock  scheme  \cite{sondhi1,indiana,fertig}.  For slowly varying
textures the two  approaches agree.

In this letter we consider textures in a more general setting, which
allows for  partially  polarized states, and which we hope will be
useful in the study of textured edges and of the response to slowly
varying  external fields.

Our starting point is a  Chern-Simons-Landau-Ginzburg (CSLG) lagrangian
for a two component system. This theory has two CS fields, which are
needed to describe partially polarized states, and an integer  valued
coupling matrix that determines filling fractions as well as charges and
statistics of the quasiparticles. The resulting effective long-distance
spin theory turns out to be a nonlinear sigma model covariantly coupled
to a CS field, which survives from the original CSLG theory.

The presence of a gauge field in the sigma model allows for non-singular
finite energy solutions of skyrmion type. The (topological) charge of
the  skyrmions is quantized, just as in the usual sigma model, but the
coupling to the CS field along with the requirement of finite energy
also quantizes the $z$-component of the spin.\footnote{  Technically
this is possible since the coupling of the CS field breaks
scale-invariance even in the absence of Coloumb and Zeeman terms, so the
skyrmions can never become large. If one gives up the requirement that
the  energy of the excitations is finite then they can be large, with a
large  (unquantized) spin.}  This is important, since the vortex
solutions in the original CSLG model have quantized charge and spin, and
we should emphasize that in the ordinary sigma model the spin is not
quantized.  We thus believe that the covariant sigma model provides the
correct spin theory for the partially polarized states. It is another
question  whether or not the skyrmions  in the long distance theory give
a good quantitative descriptions of the original vortices. Only
numerical calculations can answer this.

It turns out that the fully polarized case is special; the spin is not
quantized,  and the size of the skyrmions is determined by the Coulomb
and Zeeman terms, just as in the usual sigma model aproach.  We note
that CSLG theories with two CS fields have  been discussed before
\cite{wenzee,ezawa,indiana}, and those of our results that are purely
kinematical, such as filling fractions and quantum numbers,  are
certainly not new.

We consider a two dimensional electron gas subject to a magnetic field
perpendicular to the plane of motion. Following the general strategy for
mean field calculations, we start from the following lagrangian
describing the electrons in terms of a two-component {\em bosonic} field
$\phi$,
\be
{\cal L} &=& \phidag (i\pa 0 - \aeo - \ato\sz - A_0)\phi
 -\kappa|(i\vec \nabla +\aev + \atv\sz + \vec A)\phi|^2 \nonumber \\
  & -&\frac 1 {\pi} \ki \alpha \beta a_\alpha^0b_\beta
-\half \mu B\phidag\sz\phi - V(\phidag\phi) \ \ \ \ \ .
\label{1}
\ee
The two (Coulomb gauge) CS gauge potentials
$a^\mu_1$ and $a^\mu_2$  couple to the charge and the $z$-component of
the spin densities respectively. This  Chern-Simons lagrangian\footnote{
We shall consistently use Coulomb gauge, but it is straightforward to
introduce (redundant) longitudinal parts of the CS fields to obtain  the
usual form $\sim \ki \alpha \beta \epsilon_{\mu\nu\rho}
a_\alpha^\mu\partial^\nu a_\beta^\rho$ for the CS lagrangian \cite{zhk}.
} is such that flux quanta of the magnetic fields
$b_\alpha = \epsilon_{ij}\partial^ia_\alpha^j$ are attached to the
bosons described by $\phi$ so that they effectively become fermions.
One can show that the most general way to achieve this is to take $\ki
\alpha \beta = K^{-1}_{\alpha\beta}$ where  $K_{\alpha\beta}$ is a
symmetric
$2\times 2$  matrix with integer entries whose diagonal elements are
both either even or  odd \cite{frohlich,wenreview}.
$B\hat e_z = \nabla \times \vec A$ is the external magnetic field,
$A_0$ is the external scalar potential and the ``effective mass'',
$1/2\kappa$, and the magnetic moment, $\mu$, are phenomenological
parameters.

To disentangle the charge and
spin degrees of freedom we decompose $\phi$ as,
\be
\phi = \fii \chi \ \ \ \ \ ,
\ee
where  $\chidag\chi=1$ and the real field $\fii$ is related to the
density,
$\rho$, by $\fii =\sqrt{\phidag\phi} = \sqrt{\rho}$.  The CP(1) field
$\chi$  is related to the spin (unit)vector
$\nh$ by $\nh = \chidag\vec\sigma \chi$. We also introduce the gauge
potential $\at_\mu = i\chidag\pa \mu \chi \, , \ \mu=0,1,2$.  Note that
$\at$ is not a separate dynamical field, but determined  by $\chi$. (The
degrees of freedom in $\chi$ can conveniently be thought of as the two
angles describing the direction of $\nh$ plus an additional  overall
phase.)
transformations $\chi \rightarrow e^{i\alpha(\vec x,t)}\chi$
corresponding to $a_1^\mu\rightarrow a_1^\mu + \partial^\mu\alpha$,  but
under the transformations $\chi
\rightarrow e^{i\beta(\vec x,t)\sz}\chi$, corresponding to
$a_2^\mu\rightarrow a_2^\mu + \partial^\mu\beta$,  it rotates around the
$z$-axis as $\nh  \rightarrow e^{2i\beta(x)L_z}\nh$,  where the $3
\times 3$ matrix $L_z$ is the $z$-component of the angular  momentum.
Also note that $\tilde a^\mu$ transforms as
$\tilde a^\mu\rightarrow \tilde a^\mu - \partial^\mu\alpha -
n_z\partial^\mu\beta$, so the combination
$a_1^\mu - n_za_2^\mu - \tilde a^\mu$ is gauge invariant.  It is now
merely a matter of algebraic manipulations to rewrite (1) as,
\be
{\cal L} &=& \fii(i\pa 0 +\kappa \nabla^2)\fii - V(\rho) -
\rho(A_0  +\aeo + \ato n_z + \at_0  )
-\kappa\rho(\vec A + \aev + \atv n_z + \atvidv)^2
\nonumber \\
&+& \frac \kappa 4 \rho(\vec D^{ab}\nh_b)^2 - \frac 1 {\pi}
\ki \alpha \beta a_\alpha^0b_\beta
-\half \mu B \rho n_z \ \ \ , \label{3}
\ee
where the covariant derivative is defined by $\vec D^{ab} =
\delta^{ab}\vec\nabla + 2i\atv L_z^{ab}$.  From the transformation
properties of $\at^\mu$, it is easily established that, except for the
gauge fixed CS-lagrangian, all terms in (\ref{3}) are
invariant under each of the two gauge transformations.

We now look for solutions to the equations of motion following from
$\cal L$.  Following the usual line of arguments\cite{zhk} we impose the
condition
$\vec A + \aev + \atv n_z + \atvidv = 0$, which implies that the
external  magnetic field is cancelled by the internal fields, \ie
$B+b_1+\hat b_2 +\tilde b = 0$,  where $\hat b_2 =
\epsilon_{ij}\partial^i(a_2^jn_z)$. Varying $\cal L$ with  respect to
$a^0_{\alpha}$ gives the constraints  $\pi\rho = -\ki 1 1 b_1 - \ki 1 2
b_2 $ and  $\pi\rho n_z= -\ki 1 2 b_1 - \ki 2 2 b_2 $. For vanishing
Zeeman energy ($\mu=0$)  one easily verifies that a solution to the
equations of motion is:
$\rho = \ki 1 1 B/\pi \equiv \rhobar$ (where $V'(\rhobar)=0$),
$a^0_\alpha = a_2^\mu = \hat a_2^\mu=
\tilde a^\mu = 0$, and the spin  vector is an arbitrary constant unit
vector with fixed $z$-component
$n_z= \frac {\ki 1 2 }{\ki 1 1 }$.  This is a quantum Hall  state with
filling fraction, $\nu=2\pi \bar \rho /B$, and polarization,
$\bar n_z$,
\be
\nu = 2\ki 1 1 = \frac {2 \km 2 2} {\km 1 1 \km 2 2 - \km 1 2 ^2} \ \ ,
 \  \ \ \ \bar n_z \equiv \frac {\ki 1 2 }{\ki 1 1 } = - \frac {\km 1 2}
{\km 2 2} \ \ .
\label{mfv}
\ee
The corresponding filling fractions for the spin up and
spin down states are $\nu_\uparrow = \ki 1 1 +\ki 1 2$ and
$\nu_\downarrow = \ki 1 1 - \ki 1 2$, as discussed in \eg\cite{ezawa}.

When the Zeeman energy is included, the solution above is modified.
Instead  of pointing in a fixed direction, the (still $\vec
x$-independent) spin  vector $\hat n$ precesses around the magnetic
field; the filling fraction and  polarization still being given by
(\ref{mfv}). Note that these states are  partially polarized.

We now derive our final form of the low energy lagrangian.
Using $b_1 = -(B+\hat b_2 + \tilde b)$, and making the same type of
approximation as in \cite{lee,sondhi1,indiana},
namely  neglecting terms $\sim\partial_0\fii$ and
$\sim\vec\nabla\fii$,
we are left with
the following lagrangian (still in Coulomb gauge) for the fields $\rho$,
$\nh$ and
$a_2^\mu$:
\be {\cal L} &=& \rho\left[\at^0 + \frac  \kappa 4 (\vec D^{ab}\nh_b)^2
     -\half \mu B n_z - A^0\right] - V(\rho) \nonumber \\
&-& 2\ato \delta \rho_s - \frac 1
\pi \ato[\ki 2 2 b_2 - \ki 1 2 (\hat b_2 + \tilde b)] \ \ \ \ \ .
\label{7}
\ee
Here,
\be
\rho &=&  \rhobar + \delta\rho = \rhobar + \frac 1 \pi [ \ki 1 1 (\hat
b_2 +
\tilde b) - \ki 1 2 b_2] \nonumber \\
\rho n_z  &=& \bar\rho\bar n_z + 2\delta \rho_s \ \ \ \ \ , \label{5}
\ee
is the charge and spin density respectively, with the ground state
values, $\rhobar$ and
$\bar n_z$ given above.  In deriving (\ref{7}) from (\ref{3}) we
integrated out $a^0_1$, but kept $a^0_2$.   Note that although $\hat
b_2$ and $\tilde b$ are not separately invariant under the  remaining
gauge transformation, the combination $\hat b_2 + \tilde b$
is.\footnote{Also note that this combination is the curl of the
covariant version of
$\tilde a^\mu$, given by $\bar a^\mu = i\chidag D^\mu \chi =
i\chidag(\partial^\mu -i\sigma^z a_2^\mu)\chi= \at^\mu + n_z a_2^\mu
$.      } The mean field solution is regained from (\ref{7}) by first
minimizing the  spin-stiffness term by taking $\nh$ constant and $\vec
a_2 = 0$, this implies
$\hat b_2 = 0$ and minimizing $V(\rho)$ then also gives $\vec {\tilde a}
= 0$, and,  solving the $a_2^0$ constraint, $\delta \rho_z = 0$.

The lagrangian (\ref{7}) contains terms
$\sim \delta\rho(\vec D^{ab}\nh_b)^2$ and $\sim
\ato\rhobar (n_z-\bar n_z)$ that break scale invariance. The first is
usually neglected since it is higher in derivatives of $\nh$ but in our
case this might not be allowed since the skyrmions will be rather small.
The second term depends explicitly on the CS field and is important for
determining the size of the quasiparticles. Note that since the spin is
quantized, the Zeeman term will be a constant independent of the profile
of the skyrmion.

Before we study quasiparticle solutions, we comment on the special case
of a fully polarized state. One can then
 use the lagrangian of Lee and Kane \cite{lee} with only one CS field.
This is obtained from (1) by letting $a_{2\mu}=0$. All manipulations
leading to (\ref{7}) go through as before and the final result is
obtained  by letting $a_2^0 =0$ in (\ref{7}). This gives the lagrangian
used by Sondhi {\em et.al.}, if we use that $\at_0 = \frac 1 2 {\cal
A}^a(\nh)\partial_0\nh^a$ (up to a total time  derivative), where
${\cal A}^a$ is the vector potential of a unit monopole, and note that,
\be
\delta\rho = \frac 1 \pi \ki 1 1 \tilde b = 2\ki 1 1 \tilde\rho
\ee
where $\tilde\rho$ is the Pontryagin density,
$8\pi\tilde\rho=\epsilon^{ij}\nh\cdot(\partial_i\nh\times\partial_j\nh)$.
Note that (\ref{7}) contains the (Coulomb gauge) Hopf term,\footnote{To
get the covariant form of the Hopf term, one must first introduce the
longitudinal parts of the CS field to get the term $\frac {\ki 1 1}
{2\pi} \epsilon_{ij} a_1^i\partial_0 a_1^j$, and then note that (for
$a_2^\mu = 0$), the mean field condition implies
$\partial_0 a_1^i =-\partial_0 \tilde a^i$.   }
\be
{\cal L}_H = \delta\rho \at_0  = \frac 1 \pi \ki 1 1 \at_0\tilde b
\ee
which is needed to give the correct statistics to the quasiparticles.
That such a Hopf term should be present in the effective lagrangian was
proposed earlier, but to our knowledge it has not previously been
derived. (The Hopf  term is usually ignored since it is high in
derivatives.)

When discussing the skyrmion type quasiparticles, we must treat the
fully polarized case separately. We start with the partially polarized
case where $|n_z|<1$, and  consider a general, static, rotationally
symmetric, vortex solution:
\be
\chi(\vec r) = \left( \begin{array}{c} \cos\frac \ar 2 e^{in\phi} \\
     \sin \frac \ar 2 e^{im\phi}   \end{array} \right) \ , \ \ \ \ \ \ \
\ \
\vec a_2(\vec r) =  a(r)\hat e_\phi \ , \ \ \ \ \ \ \ \ato(\vec r) =
a^0(r) \ \ \ , \label{10}
\ee
where $(r,\phi)$ are polar coordinates. This ansatz for $\chi$ implies
\be
\nh = (\sin\ar \cos[(m-n)\phi],\, \sin\ar \sin[(m-n)\phi],\, \cos\ar) \
\ \ \ \ ,
\ee
and $\atvidv  = -(n\cos^2\frac\ar 2 + m\sin^2\frac\ar2 )\frac {\hat
e_\phi} r$.   It is straightforward to substitute  (\ref{10}) in the
e.o.m. derived from the lagrangian (\ref{7}) to get three  differential
equations for the functions $\ar$, $a(r)$ and $a^0(r)$. The detailed
expressions will be presented elsewhere, and here we shall only discuss
some general properties of the solutions. First we notice that only those
with either $n$ or $m$ equal to zero are consistent with a constant
$\varphi = \sqrt{\rhobar}$. If there is vorticity in both the upper and
lower spinor component, the density must go to zero at the center of the
vortex to avoid singularities. However, if \eg $n=0$, we can avoid the
singularity by taking $\alpha(0) = 0$, while still keeping $\rho(0)$
finite. These are the smooth ``skyrmion''  solutions referred to
earlier. To see that the configuration given by (\ref{10}) can have
finite energy, it is sufficient to note that
\be
\vec D^{ab}\nh^b = \hat e_r \partial_r\nh^a + \hat e_\phi\left[ \frac
{\partial_{\phi}} r \delta^{ab} + 2ia(r)L^{ab}_z\right]\nh^b \ \ \ \ \ ,
\label{covdiv}
\ee
so if
\be
a(r) \rightarrow \frac {n-m} {2r} \ \ \ \ \ {\rm and}\ \ \ \ \ \cos\ar
\rightarrow \bar n_z  \ \ \ \ \ , \label{asym}
\ee
for $r\rightarrow \infty$,  the covariant derivative in (\ref{covdiv})
vanishes, and $n_z$ takes its asymptotic value, which are the conditions
for having finite energy. Note that the gauge potential is crucial in
order to have finite energy ``skyrmions''.  It is at this point the
logic will differ for a fully polarized state, since  in this case the
field $\atv$ decouple from $\nh$ in the ground state, and there  will be
no extra condition of the type (\ref{asym}).    In the usual sigma
model, configurations of the type (\ref{10}) are logaritmically
divergent except for $\alpha(\infty$) equal to 0 or $\pi$ corresponding
to $n_z =\pm1$, \ie a fully polarized state having the usual skyrmions.
This logarithmic divergence was discussed in
\cite{indiana}.

The spin and charge (electric charge $=-eQ$), of a quasiparticle  are
given by
\be
Q = \int d^2x\, \delta\rho\ \ \ \ \ {\rm and}\ \ \ \ \
 S= \int d^2x\, \delta \rho_s \ \ \ \ \ , \label{chsp}
\ee
Note that although the total charge  is given by   the topological
charge, just as in the case of the usual skyrmions considered in
\cite{sondhi1,indiana}, the charge density $\delta \rho$ in (\ref{5})
is no longer given simply by the Pontryagin density $\tilde b/2\pi$.

Combining (\ref{5}) and (\ref{chsp}) with the asymptotic values in
(\ref{asym}), it is easy to derive,
\be
Q =  \frac {\ki 1 1 } \pi \int d^2x\, \tilde b =  -\frac {\nu} 2
(1+\bar n_z)n - \frac {\nu} 2 (1-\bar n_z)m \ \ \ \ \ ,
\ee
so the quasiparticle charge is determined by the filling fraction
and the polarization, or equivalently, by the filling fractions of the
two spin levels.  Similarly, by combining the $a_2^0$ constraint
equation from (\ref{5}) with the asymptotic condition (\ref{asym}), we
get  the spin of the quasiparticle as
\be
S =-\frac 1 {2\pi} \int d^2x\, [\ki 2 2 b_2 - \ki 1 2 (\hat b_2 +
\tilde b)] = - \frac 1 4 \left(\nu \bar n_z+ 2 \ki 2 2 \right)n
  - \frac 1 4 \left(\nu \bar n_z - 2 \ki 2 2 \right)m \ \ \ \ \ ,
\label{svalue}
\ee
which depends on the extra parameter $\ki 2 2$.  Note that  both
charge and spin of the quasi particles are quantized, a result that was
obtained earlier \cite{ezawa}.  This is a consequence of having two U(1)
CS fields and requiring the quasiparticle energy to be finite.

The fully polarized case, $n_z=\pm 1$, needs special attention since, as
we have already mentioned, the finite energy condition then does not
determine the asymptotic form of $\atv$, and thus leaves the  $b_2$ flux
undetermined. If we assume the condition $n_z=\frac{l_{12}}{l_{11}}$ to
be satisfied as in the partially polarized case, there is now a
restriction on the parameters,
$l_{12}=\pm l_{11}$. The filling fraction is one over an odd integer,
just as in the mean field description with a single (odd integer)
statistics parameter. Furthermore, the quasiparticle charge
$Q$ in this case is quantized whereas the spin $S$ is instead determined
dynamically by the Zeeman term. It is pleasing that our generalized
sigma model description of the fully polarized case does  not
qualitatively differ from the one given by the standard sigma model,
when it comes to  the quasi particles. It is however not excluded that
other properties, like correlation functions, and edge excitations, will
be qualitatively different.

However, there are also fully polarized ground states where the condition
$l_{12}=\pm l_{11}$ is not satisfied. These are somewhat different. In
these cases neither the charge nor the spin of the quasiparticles are
quantized. Instead a linear combination of these, with coefficients
determined by the matrices
$K_{\alpha \beta}$, will be identical to the (integer) topological
charge of the $\vec n$-field.

We end with a few comments:\\
1. A separate analysis is needed to check the stability of the mean
field ground states. For sufficiently large Zeeman coupling one would
expect the states which are not fully polarized in the direction of the
magnetic field to become unstable. It is of interest to examine the
conditions for stability relative to small oscillations as well as
relative to quasiparticle creation. \\
2. One of the motivations for this work was to find a mean field theory
for the edge excitations. It is believed that the number and
properties   of the (gapless) edge excitations can be inferred from the
properties of the bulk  state \cite{wenreview}. In particular, in a
(abelian) CS description (related to the one employed here via a duality
transformation\cite{dualwe,dualthey}) a simple gauge argument due to Wen
shows that there are equally many edge modes as CS fields
\cite{wenreview}. It has recently been shown  that the ground state at
$\nu=\bar n_z=1$ has a spin texture of the skyrmion type along  the edge
for suitable strength of the confining potential, and it is proposed
that there is a related gapless excitation, in addition to the usual
gapless density wave \cite{karlhede}.   It is thus rather natural that
the effective CS theory should contain two gauge fields,  and in this
context it would be interesting to study the  dual CS theory
corresponding   to (\ref{7}). \\
3. An important question which we have
not addressed in this paper concerns the collective modes. For the
polarized state with $n_z=\frac{l_{12}}{l_{11}}$, we expect a  spin wave
with a gap given by the Zeeman energy, just as in the usual sigma model
description. For the partially polarized states we expect a gap for the
spin wave proportional to the cyclotron energy. As discussed in
\cite{indiana}, this can not  be correct since there are empty states in
the lowest Landau level so the gap should be determined by the Zeeman
and Coulomb energies. We have no resolution to this puzzle.\\

\acknowledgements We thank Dan Arovas and Shou-Cheng Zhang for
discussions.  AK was partially supported by the Swedish Natural Science
Research Council. THH thanks NorFA for financial support to visits at
the Centre for Advanced Study.

\end{document}